# Controlled Natural Languages for Specifying Business Intelligence Applications

*Pedro das Neves Rodrigues, Alberto Rodrigues da Silva*
*INESC-ID, Instituto Superior Técnico, Universidade de Lisboa, Lisboa, Portugal*

## Abstract

This study examines the use of controlled natural languages (CNLs) to specify business intelligence (BI) application requirements. Two varieties of CNLs, CNL-BI and ITLingo ASL (ASL), were employed. A hypothetical BI application, MEDBuddy-BI, was developed for the National Health Service (NHS) to demonstrate how the languages can be used. MEDBuddy-BI leverages patient data, including interactions and appointments, to improve healthcare services. The research outlines the application of CNL-BI and ASL in BI. It details how these languages effectively describe complex data, user interfaces, and various BI application functions. Using the MEDBuddy-BI running example.

**Keywords:** Controlled Natural Language, Business Intelligence, Requirements Engineering, Data Analytics, Microsoft PowerBI

## 1. Introduction

Developing effective information systems (IS) shall involve a rigorous and systematic requirements process, a stage often compromised by the intrinsic ambiguity of natural language [1, 2]. Inadequacy of such requirements impacts financial costs, project timelines and stakeholder satisfaction. The CHAOS Report presented by The Standish Group [3] revealed that only about 40% of projects are completed on time and within budget, with inadequate requirements as a significant factor. Furthermore, unclear, and ambiguous requirements frequently lead to stakeholder dissatisfaction, reducing trust and potentially derailing projects [4].

In IS, formal languages offer precision, demand specialised training and are often less applicable to non-critical systems [5, 6]. As a promising alternative, CNLs have emerged blending the accessibility of natural language with the precision of formal languages [2].

This research focuses on applying two controlled natural languages (CNL) - i.e., CNL-BI and ITLingo ASL (ASL) for the requirements specification of BI applications. The main purpose was to evaluate the efficacy of CNL-BI and ASL in mitigating the identified risks by enhancing the quality of these requirements. An illustrative example of the MEDBuddy-BI supports this work. This fictitious BI application uses patient interactions and appointment scheduling data to improve the National Health Service's (NHS) efficiency. Using these CNLs, the MEDBuddy-BI example shows empirical insights into its effectiveness, particularly in dealing with complex domains involving multidimensional data models, dashboards designed with interactive user interface (UI) elements and use case specification.

A survey was designed to evaluate the utility and usability of CNL-BI and ASL for describing BI applications.

## 2. Controlled Natural Languages

Natural language-based requirements specifications provide vivid and easily understandable information, although their imprecision owing to unavoidable ambiguities and inconsistencies can make them challenging for automated computer processing [7]. Moreover, formal language methods can reduce errors caused by vagueness, but they might not be suitable for non-critical applications due to their specialised training and time requirements [8]. To eliminate the disparity between the ease of understanding natural language and the exactitude of formal language, CNLs were developed [2]. This language type confines the structure and vocabulary of natural language to boost clarity and support computational controllability [9, 10].

The benefits of using CNLs include (a) easier comprehension, as they closely resemble natural language; (b) reduced ambiguity, simplified grammar, and well-defined lexicon; and (c) a machine-readable syntax, eased by formal grammar and predefined terminology [2]. This research explores two linguistic styles: an informal defined CNL, named CNL-BI, and a formally defined CNL, named ASL language.

## 2.1. Linguistic Patterns

As da Silva [11] has previously discussed, a linguistic pattern consists of rules establishing the elements and terminology used in technical requirement documents. An element rule is defined by a series of attributes such as <id>, <name>, or <type>, each of which is determined by properties such as name, type, and multiplicity. In addition, a vocabulary guideline defines a compilation of specific terms such as "User" "ExternalSystem" or "Timer" that categorise certain element attributes and restrict the usage to a specific set of terms. For instance, the linguistic pattern "Actor" is determined by a set of rules, specifically the "Actor" element rule and the "ActorType" vocabulary rule specified above (cf. Spec 1).

```
Actor::
    <id:ID> <name:String> <type:ActorType>
    <stakeholder:Stakeholder>? <isA:Actor>
    <description:String>?

ActorType::User | ExternalSystem
```

*Spec 1 - Actor's linguistic pattern example (obtained from the work of da Silva et al. [12])*

In the example Spec 1, previously described by da Silva et al. [1, 12], the "Actor" element rule delineates specific attributes, including "id", "name", and "type". These attributes are each further defined by their respective name, data type and multiplicity. The multiplicity factor may be mandatory ("1" representing the default), optional ("?"), singular or plural ("+"), or non-existent ("*"). Attributes can take on different types, such as basic data types like "ID" and "String", element types, or vocabulary types like "ActorType". Moreover, the "ActorType vocabulary rule" is indicated with the "enum" tag and confines its values to literals, notably "User" and "ExternalSystem". In essence, a linguistic pattern in this key area consists of two principal elements: (1) The system comprises a collection of element attributes, each being assigned a name, type, and multiplicity; (2) The system utilises a constrained vocabulary consisting of a finite set of terms.

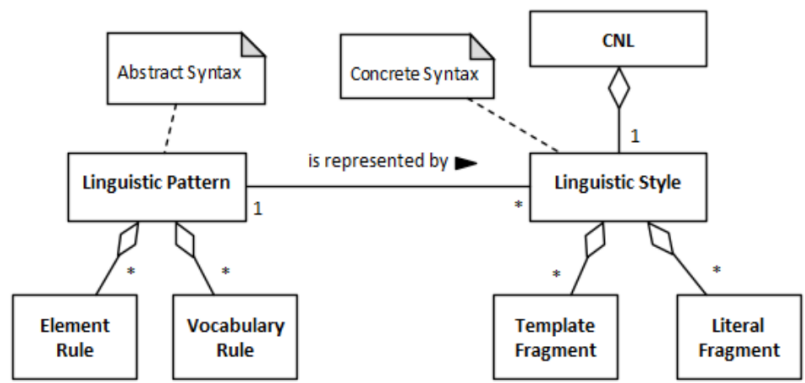

*Figure 1 - Relation between linguistic pattern and linguistic style (Obtained from the work of da Silva et al. [12])*

As shown in Figure 1, sourced from da Silva et al. [12] research, a linguistic pattern is not limited to one form but can be expressed in multiple ways to suit users' specific requirements and goals. Within this context, a "linguistic style" is established as a distinct manifestation of a linguistic pattern. This signifies that a linguistic style acts as a designated template where the characteristics of the fundamental linguistic structure can be replaced. Hence, one linguistic structure can be showcased through various linguistic styles, such as CNL-BI and ASL.

## 2.2. Linguistic Styles

A linguistic style consists of two types of text segments arranged sequentially: fixed text fragments such as "actor", "[", "]", or ",", and variable template fragments, which are denoted by the format "<element_name.attribute_name>". In this format, the element name is followed by its attribute name and enclosed within angle brackets (e.g. "<actor.type>"). Symbols like "?", "+" and "*" indicate the multiplicity constraints.

This format consistently represents the CNL-BI linguistic style throughout this paper, simplifying its presentation (cf. Spec 2).

```
Actor <id> "<name>"? is a <type> [, extends <isA>]?, with stakeholder <stakeholder>]?, described as
<description>]?.
```
*Spec 2 - Linguistic style of the Actor element in CNL-BI*

We used the representation as represented in Spec 3 to define ASL linguistic style.

```
Actor <id> "<name>"? : <type> [
    (isA <isA>)?
    (stakeholder <stakeholder>)
    (description <description>)?
]
```
*Spec 3 - Linguistic style of the Actor element in ASL*

## 2.3. CNL-BI

This study proposes the CNL-BI language, inspired by the work previously discussed by da Silva and Savic [1, 12]. CNL-BI is defined as specifying BI applications based on linguistic patterns to write requirements better. CNL-BI inherently supports various concepts, including data entities, data attributes, primitive types, and enumerations, as defined previously by da Silva and Savic [1], but also linguistic patterns to represent BI-specific concepts, such as data entities in multidimensional models, Online Analytical Processing (OLAP) operations, and user interface (UI) graphical elements, such as interactive and responsive charts. An instance of this is seen in Spec 4, which presents a representation of an "Institution" using CNL-BI.

```
DataEntity Institution is a Master Dimension with attributes
    id is a UUID (PrimaryKey),
    code is a String (NotNull),
    name is a String (NotNull),
    latitude is an Integer (NotNull),
    longitude is an Integer (NotNull),
    location is a String (NotNull),
    type is a String (NotNull)
described as this dimension represents the details of an institution.
```
*Spec 4 - CNL-BI representation of an Institution dimension data entity*

## 2.4. ASL Language

ITLingo ASL (ASL) [13] is a specification language used to outline software business applications and is an essential component of the ITLingo Initiative [14]. ASL integrates principles from ITLingo RSL (RSL) [1] and Object Management Group's Interaction Flow Modelling Language (IFML) [15]. ITLingo RSL is a language for specifying requirements and testing [16, 1, 11, 17]. In contrast, IFML is a modelling language employed to define the user interface components of an application's front end [15].

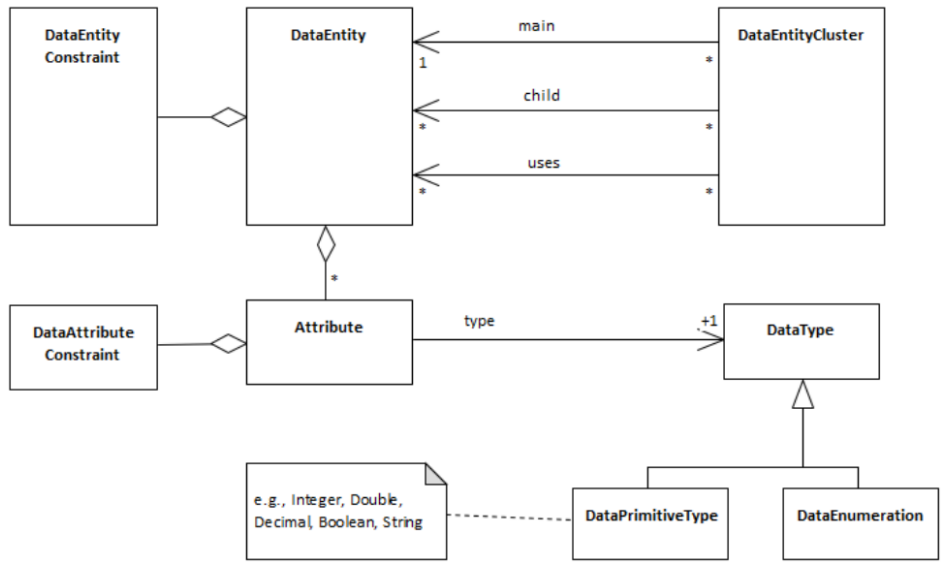

*Figure 2 - ASL partial DataEntities metamodel with data entities related constructs (UML notation). Obtained from the work of da Silva et al. [1]*

The key concepts outlined by ASL in this research are (1) data entities and enumerations, which represent domain-specific ideas, displayed in Figure 2; (2) UI elements, which include UI containers, components, and parts; (3) actors, which indicate the roles executed by users or other systems; and (4) use cases, which describe interactions between the actor(s) and the system being investigated. These concepts are systematically arranged into two primary architectural views: (1) the data view and the use case view (both based on the RSL [1] language) and (2) the user interface view (using IFML [15] language as inspiration).

The same example of a Health Institution can be represented using ASL as defined in Spec 5.

```
DataEntity Institution "Institution" : Master [
    attribute id "UUID" : UUID [constraints (PrimaryKey NotNull Unique)]
    attribute Code "Code" : String(50) [constraints (NotNull)]
    attribute Name "Name" : String(50) [constraints (NotNull)]
    attribute latitude "Latitude" : Integer [constraints (NotNull)]
    attribute longitude "Longitude" : Integer [constraints (NotNull)]
    attribute location "Location" : String(100) [constraints (NotNull)]
    attribute type "Type" : String(50) [constraints (NotNull)]
    description "This dimension represents the details of an institution" ]
```

*Spec 5 - ASL representation of an Institution*

## 3. Multidimensional Analytics

Multi-perspective data analysis using multidimensional techniques is a powerful way to uncover patterns and insights, particularly with large datasets [18]. It is an invaluable technique for data-driven decision-making. Data warehouses use specific structures, with fact tables attached to dimension tables, typically organised into star or snowflake schemas [19]. With a star schema, dimension tables connect directly to a central fact table. This configuration streamlines and accelerates data queries due to its simplicity. In contrast, a snowflake schema partitions dimension tables into smaller fragments, emphasising data organisation and precision [19].

Such structures are transformed into OLAP cubes when they become excessively intricate for speedy data queries. These cubes resemble three-dimensional data environments with the fact table as the primary aspect and dimensions on the other sides [20]. OLAP enables various techniques for analysing data using five main operations: slice, dice, roll-up, drill-down, and pivot [20]. These actions are illustrated in Figure 3.

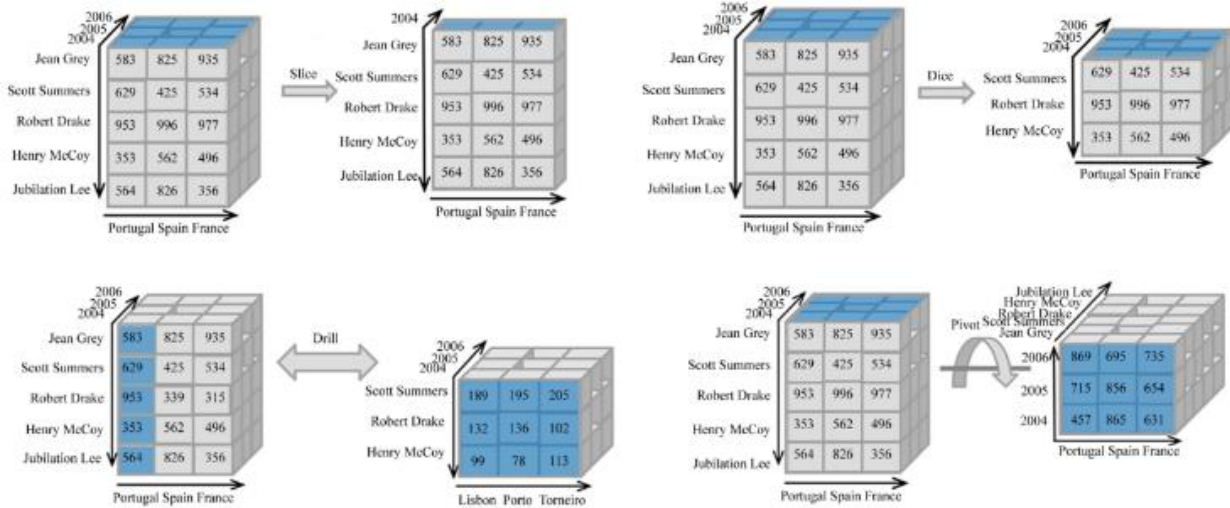

*Figure 3 - Representation of a cube operations: slice (top-left), dice (top-right), roll-up/drill-down (bottom-left) and pivot (bottom-right). Obtained from the work of Ribeiro et al. [20].*

A unique query language, distinct from Structured Query Language (SQL), is typically used for big data in OLAP. This language, such as Multidimensional Expressions (MDX) [21], facilitates intricate data analysis and efficient management of vast data sets. Despite commonly perceiving OLAP cubes as three dimensions, they can handle even more complex, multidimensional data sets. To summarise, OLAP cubes and data warehouses are vital instruments for advanced data analysis, providing significantly more benefits than conventional database systems [22].

## 4. Illustrative Example: MEDBuddy-BI

MEDBuddy was designed as a digital intermediary to enhance access to healthcare services within the NHS. It offers a straightforward method for patients to manage their medical appointments. The scheduling function of the app is integrated with real-time information from healthcare providers, enhancing appointment booking efficiency and reducing administrative workloads. Automated reminders are intended to decrease the number of missed appointments, thus optimising the utilisation of NHS resources. Moreover, MEDBuddy's centralised medical appointment record aims to empower patients by providing a comprehensive history of their healthcare interactions, facilitating greater patient involvement in their health management.

MEDBuddy-BI is an analytical add-on to the MEDBuddy application that aims to enhance efficiency and patient care within the NHS by drawing on data collected from patient interactions and appointment scheduling [23]. The tool is equipped with an analytics engine for real-time data processing, which then translates the gathered information into easy-to-use dashboards and reports for healthcare administrators and policymakers to scrutinise trend analyses, locate shortcomings, and predict future service demands.

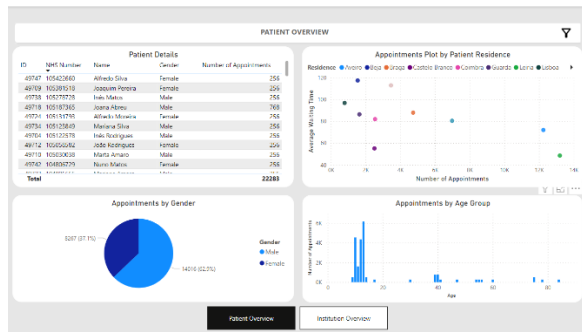

*Figure 4 – MEDBuddy-BI Patient Overview Page*

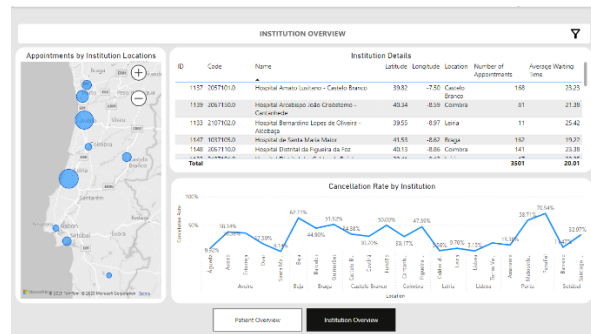

*Figure 5- MEDBuddy-BI Institution Overview Page*

The BI tool has advanced data visualisation features that enable quick decision-making, detailed feedback analysis that enhances service quality, and strict adherence to data protection laws. It is a strategically valuable resource for NHS to become more data-focused, leading to patient care enhancements and system-wide healthcare improvements. MEDBuddy-BI offers two dashboard views: (1) a patient-focused dashboard (cf. Figure 4) that concentrates on individual patient data and yields information about healthcare usage across various demographics, such as gender, age, and residence. The dashboard's interactivity enables precise analysis that could influence health promotion and resource allocation; (2) an institution-centred dashboard (cf. Figure 5) provides a panoramic view of healthcare service availability and usage. It allows service demand comparison across regions, monitors institution cancellation rates, and profiles institutional performance.

The application is currently available for access at [23], and the comprehensive ASL description of MEDBuddy-BI can be found on this GitHub repository [24].

## 5. Data Specification

Data entities are structural components in information systems, often shaped by requirements gathering and domain evaluation activities [19, 18]. In BI applications, these data entities are integral to constructing multidimensional models, which form the foundation for various BI functionalities like data warehousing, analytics, and reporting [20].

According to Microsoft's classification [25], data entities can be categorised into different types, namely:

- **Master Data Entities** hold information central to the business and rarely changed, often serving as the backbone for transactional data.
- **Transactional Data Entities** contain business activity records subject to frequent changes and are often the focus of BI queries.
- **Reference Data Entities** hold static data for categorisation or hierarchies, such as lists of states or countries.

In a multidimensional model, these data entities often manifest as dimensions and facts that enable complex queries and data analysis [19, 18]. Specifically, fact tables are best categorised as transactional data entities containing quantitative metrics like sales, revenue, or performance indicators that change frequently. Conversely, dimension tables can be categorised as either master or reference data entities, depending on their content. Dimensions could encompass various business aspects like time, geography, or customer demographics [20]. Table 1 summarizes the fact and dimension types in the context of BI applications.

*Table 1 - DataEntity types in the context of BI applications*

| Type | Description |
|------|-------------|
| Fact | Quantifiable data points, such as sales, revenue, or profit, are typically the focus of analysis. Facts are aggregated or measured across various dimensions to generate insights. |
| Dimension | Categorical data that provides context to facts, such as time, geography, or customer demographics. Dimensions are used to slice and dice the facts, enabling a more detailed or broad analysis. |

The multidimensional model facilitates the aggregation and disaggregation of these measures across different dimensions, offering a detailed and comprehensive view of business operations. Therefore, accurate identification and categorisation of data entities are crucial for the foundational architecture of an information system and vital for enhancing BI applications' analytical capabilities.

This study uses a running example to aid the explanation and discussion of the requirements of a specialised BI application known as MEDBuddy-BI, introduced in Section 4. Figure 6 illustrates this example's primary facts and dimensions using the UML notation.

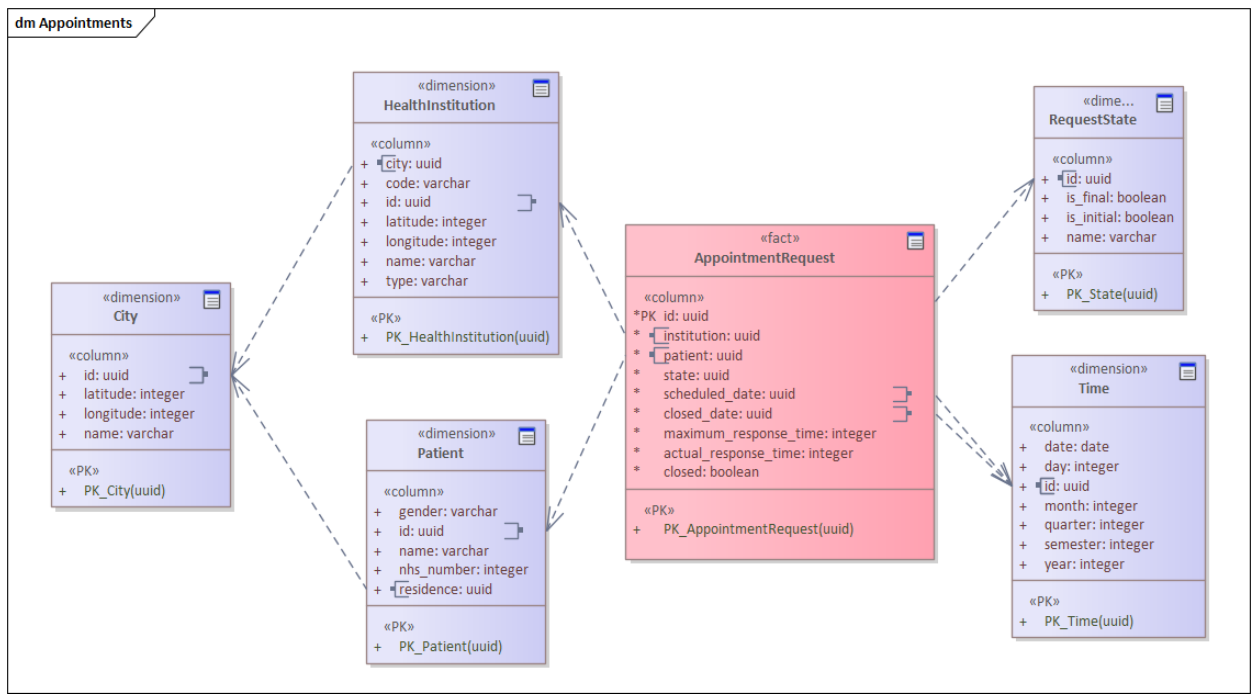

*Figure 6 – MEDBuddy-BI data model that supports the analysis of appointment requests.*

## 5.1. Linguistic Pattern

The linguistic patterns of "DataEntity" and "DataAttributeConstraint" are defined by the rules in Spec 6. Concurrently, Table 1 summarizes the fact and dimension types in the context of BI applications.
Table 1 summarises the potential subtypes that can be employed to classify a "DataEntity" during the design of a multidimensional model.

```
DataEntity::
 <id:ID> <name:String>? <type:DataEntityType> <subtype:DataEntitySubType>?
 <dataAttribute:DataAttribute>
 <dataEntityConstraint:DataAttributeConstraint>?
 <description:String>?

DataEntityType::Reference | Master | Transaction

DataEntitySubType::Fact | Dimension

DataAttribute::
 <id:ID> <name:String>? <type:DataType>
 <defaultValue:String>?
 <operation:Operation>?
 <constraints:DataAttributeConstraint>?

DataType::DataPrimitiveType | DataEnumeration

DataPrimitiveType::Integer | String | Boolean | Date...

DataEnumeration::<id:ID> <name:String>? <values:String>*
```
*Spec 6 - Linguistic Pattern of DataEntities*

When an attribute of data is joined with an associated operation, it is classified as a measure. In multidimensional models, measures are traits accessible by completing logical or arithmetical operations. In CNL-BI and ASL, the syntax for representing these operations is like other expression languages, like the one found in Microsoft PowerBI Data Analysis Expressions (DAX) language [26, 27].

## 5.2. Linguistic Styles

The statement in Spec 7 defines concrete representations for the "DataEntity" pattern according to the CNL-BI and ASL languages.

```
// CNL-BI Representation
DataEntity <id> ("<name>")? is a <type> <subtype>, with attributes <dataAttribute>+, (described as
<description>)?.
// ASL Representation
DataEntity <id> ("<name>")? : <type> : <subtype> [
    attribute <dataAttribute>+
    (description <description>)?.]
```
*Spec 7 - CNL-BI and ASL linguistic styles of a DataEntity*

## 5.3. Examples

Considering the MEDBuddy-BI example (cf. Page 17 for a complete application description), we identify and annotate the **Entities** in **bold** text.

A **request** for a medical appointment has a **patient** associated with it, the **institution** where the appointment takes place, the **date** the appointment was booked and the **date** it was completed, and the **status** of the request, which can be "booked", "held" or "cancelled", the last two being final statuses. The completion **date** is the date the appointment was made, if the appointment was held, or the date the **request** was cancelled, if the request was cancelled.
A **request** for a medical appointment also limits the number of days the appointment must take place after the request enters the "booked" status, known as the maximum response time. It also has associated with it the number of days that have elapsed between the request moving from the "booked" state to one of the completion states ("held" or "cancelled"), referred to as the actual response time.

**Patients** are characterised by their NHS identification number, gender, age, name, and residence.

A **healthcare institution** is characterised by its code, name, latitude, longitude, city, and type of institution. A health institution can be of the Health Centre or Hospital type. A **city** is characterised by its name, latitude, and longitude.

The **dates** on which a medical appointment was booked and when it occurred are described by the day, month, quarter, semester, and year in which they were recorded. This information is especially relevant for analysing metrics relating to requests for appointments in specific periods, such as the total number of appointments nationwide in 2023.

*Spec 8 - Informal description of the MEDBuddy-BI entities*

Using the multidimensional model framework, we classify each of these data entities according to a specific taxonomy, as described in  Table 1 summarizes the fact and dimension types in the context of BI applications.

Table 1. Drawing on insights obtained from domain analysis, we can discern differing entity types, as shown in the follow example: the entities "Patient", "Time", and "City" come under the classification of dimension, whereas "AppointmentRequest" is classified as a fact entity, containing data on healthcare appointments. Spec 9 outlines the definition of "Patient", "Time", "City" and "AppointmentRequest" entities in CNL-BI.

It is important to highlight that dimensions can also refer to other dimensions, such as a patient's residence, which refers to a city.

```
DataEntity Patient is a Master Dimension with attributes
    id is a UUID (PrimaryKey),
    nhs_number is an Integer (NotNull),
    name is a String (NotNull),
    gender is a Gender (NotNull),
    residence refers to Dimension City (NotNull)

DataEntity City is a Reference Dimension with attributes
    id is a UUID (PrimaryKey),
    latitude is an Decimal (NotNull),
    longitude is an Decimal (NotNull),
    name is a String (NotNull).

DataEntity Time is a Reference Dimension with attributes
    id is a UUID (PrimaryKey),
    date is a Date (NotNull),
    day is an Integer (NotNull),
    month is an Integer (NotNull),
    quarter is an Integer (NotNull),
    semester is an Integer (NotNull),
    year is an Integer (NotNull).

DataEntity AppointmentRequest is a Transactional Fact with attributes
    id is a UUID (PrimaryKey),
    institution refers to Dimension Institution (NotNull),
    patient refers to Dimension Patient (NotNull),
    state refers to Dimension RequestState (NotNull),
    scheduled_date refers to Dimension Time (NotNull),
    closed_date refers to Dimension Time,
    maximum_response_time is an Integer (NotNull),
    actual_response_time is an Integer,
    closed is a Boolean (NotNull),
    CountAppointments is an Integer (operation COUNT(id)),
    CountCancelledAppointments is an Integer (operation COUNT(state = State.Cancelled)),
    CancellationRate is a Decimal (operation (CountCancelledAppointments / CountAppointments)),
    AvgWaitingTime is a Decimal (operation AVERAGE (actual_response_time)),
    MinDate is a Date (operation MIN (scheduled_date)),
    MaxDate is a Date (operation MAX (scheduled_date)).
```

*Spec 9 - CNL-BI representation of the AppointmentRequest, Patient, Time, and City entities*

## 6. Specification of BI-based Use Cases

In software development, a use case involves interactions between a system and its users to achieve a specific objective, according to Jacobson's initial concept [28]. This concept was expanded in Use-Case 2.0 by Jacobson et al. [29] to include all user interactions with a system for a distinct aim. UML guidelines [30] outline a use case as a series of system actions leading to beneficial outcomes for users. Cockburn [31], Wirfs-Brock and Schwartz [32], and Constantine and Lockwood [33, 34] have expanded on this concept in use case scenarios, which are vital for structuring software requirements in UML. Nonetheless, their practical implementation is frequently problematic [31, 29, 35].

This study employs the UML definition of use cases [30]. It centres on actors who embody user roles or external systems, engaging with the system to achieve specific objectives. Not all stakeholders can be considered actors; for instance, a "manager" may have direct involvement that differs from a "project sponsor" [31]. Although use cases are commonly presented with scenarios and steps, da Silva et al. [12] note that such intricacy is not indispensable for our intention. In BI applications, user interactions are mainly analytical and can be effectively represented utilising OLAP operations without extensive specification of scenarios and steps.

## 6.1. Linguistic Pattern

The linguistic pattern for an actor is showcased in Spec 1. Spec 10 describes the linguistic pattern of a use case. The extension to use case types and actions are highlighted in grey.

```
UseCase::
  <id:ID> <name:String>? <type:UseCaseType>
  <stakeholder:Stakeholder >
  <primaryActor:Actor>
    <supportingActors:Actor>*
  <dataEntity:DataEntity>?
  <action:UseCasesActions>+ // see [16]
  <description:String>?

UseCaseType:: EntityCreate | EntityRead | EntityUpdate | EntityDelete | [...] | BIAnalysis

UseCasesActions:: Create | Read | Update | Delete | [...] | Slice | Dice | Roll-up | Drill-down | Pivot
```
*Spec 10 - Use case linguistic pattern*

A use case could specify numerous actions that can take place within its context, some of which are typical in IS, such as user actions (e.g., searching, filtering, creating, reading, updating, deleting, printing, closing, and cancelling) [16]. In BI applications, use cases mainly involve system users performing analytical operations and queries, including filtering and exploring data, as in this study. To execute these actions, we employed OLAP operations and expanded the number of use case actions to involve "Slice", "Dice", "Roll-up", "Drill-down", and "Pivot" actions [18]. The linguistic style of OLAP operations is visible in Spec 11.

```
Slice::<id:ID> <name:String>? <whereClause:WhereClause> <description:String>?

Dice::<id:ID> <name:String>? <whereClause:WhereClause>+ <description:String>?

WhereClause::<dimension:Dimension> | <literal:Literal>

Roll-up::<id:ID> <name:String>? <fromClause:FromClause> <description:String>?

Drill-down::<id:ID> <name:String>? <fromClause:FromClause> <description:String>?

FromClause::<attribute:DataAttribute>

Pivot::<id:ID> <name:String>? <swapClause:SwapClause> <description:String>?

SwapClause::<dimension:Dimension>+
```
*Spec 11 - Linguistic pattern of OLAP operations.*

## 6.2. Linguistic Styles

The statements in Spec 12 define concrete representations for the actor and the use case patterns according to the CNL-BI and ASL languages.

```
//CNL-BI Representation
Actor <id> <name> is a <type>
  with <stakeholder>?
  (described as <description>)?

UseCase <id> ("<name>")? is an <type>
  with stakeholder <stakeholder>, actor <primaryActor> (, support actor <supportingActor>)*
  data source <dataEntity>
  performs <action>+
  [...] // Scenarios
  (described as <description>)?

//ASL Representation
Actor <id> <name> : <type> [
  (stakeholder <stakeholder>)?
  (description <description>)? ]

UseCase <id> ("<name>")? : <type> [
  actorInitiates <primaryActor>
  (supportingActors <supportingActor>*)?
  action <action>+
  (description <description>)? ]
```
*Spec 12 - CNL-BI and ASL linguistic styles of an actor and a use case.*

The statements in Spec 13 describe the "Slice", "Dice", "Roll-up", "Drill-down" and "Pivot" OLAP operations, described on Section 3.

```
Olap Operation <id> ("<name>")? is a [Slice | Dice], where <whereClause>, described as <description>?
Olap Operation <id> ("<name>")? is a [Roll-up | Drill-down], group by <attribute>, described as <description>?
Olap Operation <id> ("<name>")? is a Pivot, swap <dimension> with <dimension>, described as <description>?
```
*Spec 13 - CNL-BI linguistic style of OLAP operations.*

## 6.3. Examples

Remaining with the example of MEDBuddy-BI (for further details, refer to Page 17), we can identify two roles that interact with the system: (1) a data analyst at the national level and (2) a manager of health institutions. The former is authorised to view appointment data at the national level, while the latter is only to view appointment information of their respective institution.

```
The system allows users to interact with it to conduct data analysis operations on the information relating to
requests for medical appointments. There are two types of users of the system:

1. The data scientist at a national level, is responsible for analysing the state of the SNS across the whole
country.
2. The data analyst at a health institution is responsible for analysing relevant aspects of requests for
consultations for their institution.

Both users can view information regarding medical appointment requests at the Patient and Institution level. In
contrast, an institution's data analyst will only be able to view this information at the level of their
institution.

On the page relating to health institutions, it is possible to view the general information of an institution
(its name, code, city, etc.), as well as the total number of orders, the total number of cancelled orders and
the average time waiting time in each institution, in tabular form. It is also possible to view this information
on a map of Portugal, and the cancellation rate for appointment requests by institution is displayed.
Finally, the user can filter the information on this page by location of the institution or group this information
by city.

On the patient page, it is also possible to view general patient information in a tabular or a graphical
representation, distributed by gender, age, and residence.

The user can filter information relating to patients or institutions over time on both pages.
```
*Spec 14 - Informal description of the MEDBuddy-BI app functionalities*

The text contained in Spec 14 presents few functions available to the user, including appointment request filtering based on a specified time frame and appointment request grouping according to the institution's name or city. Through textual analysis, it is possible to identify four principal use cases in which each actor analyses the Institution and Patient pages. Spec 15 illustrates the use case of Institution page analysis for both actors.

```
// Actors
Actor NationalLevelDataAnalyst is a User described as a user with permission to visualise the system at a
national level.
Actor InstitutionLevelDataAnalyst is a User described as a user with permission to visualise the system at an
institutional level.

// Use Cases
UseCase AnalysisAppointmentsInstitutionOnNationalLevel is a BI_Analysis
  actor NationalLevelDataAnalyst,
  data source AppointmentRequest,
  performs
    // Appointments in a specific year
    OLAP Operation ScheduledAppointmentsInSpecificYear is a Slice
      where Appointment.scheduled_date.year = Time.year
      described as slice the data to visualise the appointments that were scheduled in a specific year,
    // Appointments scheduled in a specific city and year
    OLAP Operation ScheduledAppointmentsBySpecificCityAndYear is a Dice
      where Institution.city = City.id and Appointment.scheduled_date.year = Time.year
      described as dice the data to visualise the appointments that were scheduled in a specific city and year,
    // Appointments by institution's city
    OLAP Operation AppointmentsByInstitutionCity "Appointments by institution's city" is a Roll-up
      group by Institution.city
      described as rolling up the data to visualise the number of appointments per institution's city,
    // Appointments by institution
    OLAP Operation AppointmentsByInstitution "Appointments by institution" is a Drill-down
      group by Institution.name
      described as rolling up the data to visualise the number of appointments per institution.
```

```
UseCase AnalysisAppointmentsInstitutionOnNationalLevel "Analysis Appointments on Institution Level" is a
BIAnalysis
  actor InstitutionLevelDataAnalyst,
  data source AppointmentRequest,
  performs
    // Appointments in a specific year
    Slice ScheduledAppointmentsInSpecificYear "Scheduled appointments in a specific year" is an OLAP operation
      where Appointment.scheduled_date.year = Time.year
      described as slice the data to visualise the appointments that were scheduled in a specific year,

  described as only displaying the appointment data at a single institution level.
```

*Spec 15 - CNL-BI examples of actors and use cases in the context of the MEDBuddy-BI app.*

In *AnalysisAppointmentsInstitutionsOnInstitutionLevel*, the user cannot perform actions on the institution dimension, since it should only visualise data related to their institution.

## 7. Specification of User Interface Elements

Enterprise software solutions depend on user interactions to carry out the data management activities that comprise the bulk of their knowledge base. UI elements, such as menus and forms, are usually created to create, read, update, and delete (CRUD) operations [36, 37].

While creating these UI elements, ASL uses terms from IFML [13], in which the UI structure is outlined via UI containers, UI components, and UI parts [15, 13]. Thus, describing application interfaces using ASL terminology becomes uncomplicated. For instance, a dashboard page featuring a single table and chart can be efficiently mapped out through ASL elements. The page will be arranged in a primary window container and classified as either a "Main Window" or "Modal Window" based on its application context. The table should take the form of a list. At the same time, the chart could be categorised as a detail or form type, with the possibility of sub-types such as "MasterDetail" or "Nested" for more intricate visualisations. The filter component, commonly in form format, could be a simple or multi-choice option tailored to meet user input requirements. Additionally, the navigation button would fall under the menu classification. An IFML outline of this dashboard page is depicted in Figure 7.

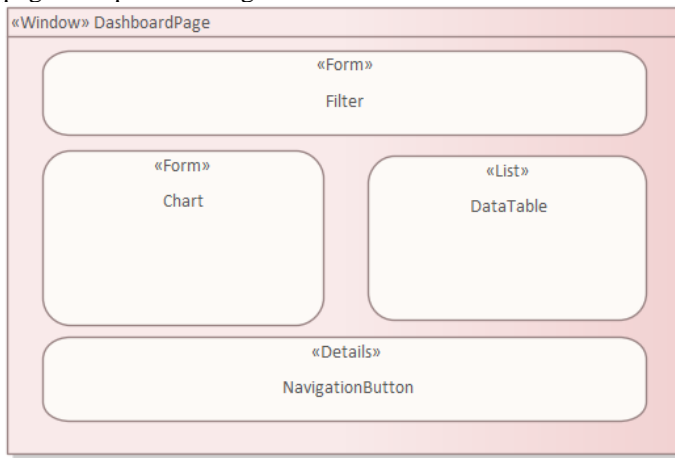

*Figure 7 - IFML representation of a dashboard page example.*

Although this vocabulary is suitable for the "Lists" and "Filters" specifications, it fails to fully comprehend the intricacy of contemporary interactive charts prevalent in applications like Power BI. We propose introducing the "InteractiveChart" component, an add-on to ASL's UI components that precisely defines modern chart interfaces.

The "InteractiveChart" component offers various chart types, such as bar, line, and pie, while delivering advanced functionalities that users or the component could activate, including drill-downs, labelling, tooltips, and real-time data updates. The component's ability to showcase dynamic data visualisations and elevate user engagement bridges a crucial gap in ASL, empowering the creation of data-driven user interfaces. A chart can assume various forms that feed to specific data visualisation requirements. For instance, it can present information through horizontal or vertical bars, where the length of the bar connotes value. This chart is known as a "BarChart". Although there are multiple chart types [38], this study focuses on a few, as summarised in Table 2.

*Table 2 - Chart subtypes (adapted from [38])*

| Type | Description |
|------|-------------|
| **Bar Chart** | Bar charts are one of the most common data visualisations. They are effective for quickly comparing data across categories, highlighting differences, showing trends and outliers, and briefly revealing historical highs and lows. |
| **Line Chart** | The line chart connects several distinct data points, presenting them as a continuous evolution. It helps view trends in data, usually over time, such as stock price changes over five years or website page views for the month. |
| **Pie Chart** | Pie charts are powerful for adding detail to other visualisations. Alone, they don't provide a quick and accurate way to compare information. They are best used to drill down on other visualisations rather than being the focus of a dashboard. |
| **Scatter Plot** | Scatter plots are adequate for investigating the relationship between different variables. They show if one variable is a good predictor of another or if they tend to change independently. The chart can be enhanced with analytics like cluster analysis or trend lines. |
| **GeographicalMap** | GeographicalMaps are ideal for visualising location information, such as postal codes, state abbreviations, or country names. They are compelling for showing how location correlates with trends in your data. |

Moreover, the chart component can be modified in response to actions initiated by the user or the application. We can draw parallels between the foundational CRUD operations and common use cases for interactive charts to describe these actions. Additionally, custom actions can be defined to align precisely with interactive chart functions, such as "DrillDown", "RealTimeDataUpdate", "ZoomAndPanUpdate", and "TooltipAndHoverDetail". Further elaboration on these actions can be found in Table 3.

*Table 3 - Possible actions to perform on an interactive chart component.*

| Action | Description |
|--------|-------------|
| **DrillDown** | This action lets users delve deeper into the chart's specific data points or categories, providing a more detailed view of the selected data segment. |
| **RealTimeDataUpdate** | This involves dynamically updating the chart as new data becomes available, ensuring the representation is always current and relevant. |
| **ZoomAndPanUpdate** | Allows users to interactively explore different areas of the chart by zooming in for detail or panning across for a broader context. |
| **TooltipAndHoverDetail** | Displays additional information, such as specific data values and insights, when the user hovers over chart elements. |

## 7.1. Linguistic Pattern

The rules in Spec 16 define the linguistic patterns of interactive chart components.

```
InteractiveChartComponent::
    <id:ID>
    <name:String>?
    <type:InteractiveChartComponentType>
    <dataBinding:DataEntity>
    <part:ChartUIPart>
    <action:InteractiveChartComponentActionTypes>
    <description:String>

InteractiveChartComponentType:: InteractiveBarChart | InteractiveLineChart | InteractivePieChart |
InteractiveGeographicalMap | InteractiveScatterPlot | ... // see table Table 2

InteractiveChartComponentActionType:: DrillDownUpdate | RealTimeDataUpdate | ZoomAndPanUpdate |
TooltipAndHoverDetailShow | ... // see table Table 3

ChartUIPart:: Axis | GeoPoint | ...
```
*Spec 16 - Interactive chart components linguistic pattern.*

## 7.2. Linguistic Styles

The statements in Spec 17 define the CNL-BI and ASL linguistic styles to represent interactive chart components.

```
// CNL-BI Representation
UIComponent <id> "<name>"? is a <type>
  data binding to <dataBinding>,
  with <part>+
  actions <action>*,
  (described as <description>)?

// ASL Representation
UIComponent <id> "<name>"? : <type> [
  dataBinding <dataBinding>
  part <part>+
  event <action>*
  (description <description>)? ]
```

*Spec 17 - CNL-BI and ASL linguistic style to represent interactive chart components.*

## 7.3. Examples

The "InstitutionOverviewPage", illustrated in Figure 8, structured as a "Main Window" container, includes key components: A form UI component named "Time Range Filter" for filtering data (number 1); a table UI component called "Institution Details" displaying multiple institution-related columns from "AppointmentRequest" data source (number 2); and an interactive line chart UI component titled "Cancellation Rate by Institution" that is data-bound to "AppointmentRequest". It highlights institution types and appointment counts while also being equipped with "DrillDown" and "RealTimeDataUpdate" actions (number 3). An interactive map interface has been designed to show "Appointments by Institution Locations". The map displays geographical data with the number of appointments shown (number 4). Navigation buttons are also available for institutional and patient page overviews (number 5).

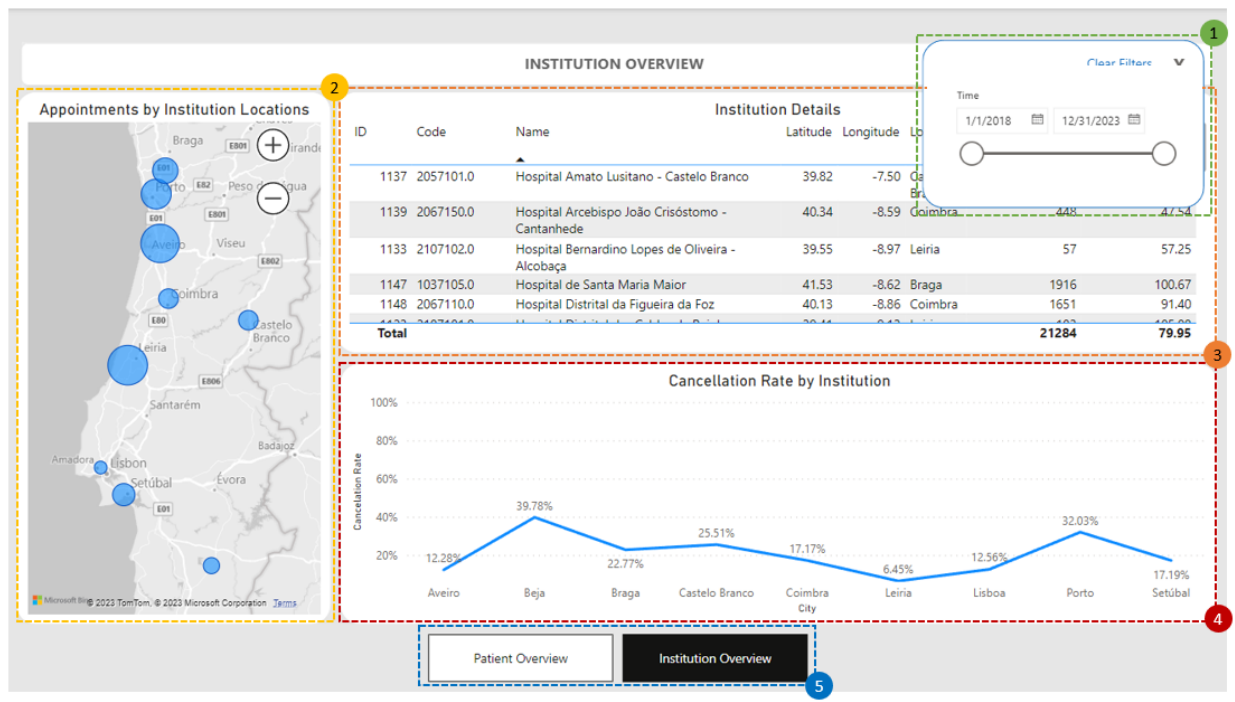

*Figure 8 - Institution page with the TimeRange (1), the LocationMap (2), the Institution Details Table (3), the Cancellation Rate Chart (4) and the NavigationButtons (5)*

The CNL-BI specification for this page is detailed in Spec 18.

```
// Institution Overview Page
UIContainer InstitutionOverviewPage is a Main Window
  that contains
    // Number 1
    UIComponent TimeRangeFilter is a Form
      data binding to AppointmentRequest,
      starting at AppointmentRequest.MinDate,
      and ending at AppointmentRequest.MaxDate,
    // Number 2
    UIComponent LocationMap is an InteractiveGeographicalMap
      data binding to AppointmentRequest
      with latitude Institution.latitude
      longitude Institution.longitude,
      and value AppointmentRequest.CountAppointments,
      actions ZoomAndPanUpdate, DrillDown, TooltipAndHoverDetails
    // Number 3
    UIComponent InstitutionTable is a Table
      data binding to AppointmentRequest
      with columns Institution.id, Institution.code, Institution.name, Institution.latitude,
Institution.longitude, Institution.location, Appointments.CountAppointments, AppointmentRequest.AvgWaitingTime,
    // Number 4
    UIComponent CancellationRateChart is an InteractiveLineChart
      data binding to AppointmentRequest
      with x-axis Intitution.type
      and y-axis AppointmentRequest.CountAppointments,
      actions DrillDown, RealTimeDataUpdate, TooltipAndHoverDetails
    // Number 5
    UIComponent InstitutionPageNavigationButton is a Detail
      that navigates to InstitutionOverviewPage,

    UIComponent PatientPageNavigationButton is a Detail
      that navigates to PatientOverviewPage.

// Patient Overview Page
UIContainer PatientOverviewPage is a Main Window [...]
```

*Spec 18 – CNL-BI representation of the Appointments Dashboard and the Institution page*

To demonstrate the dynamic and interactive properties of this page, Figure 9 presents information specific to appointments scheduled solely in 2023, following modifications to the "TimeRageFilter" and adjustments in the display of the "AppointmentsByInstitutionLocationsMap", the "InstitutionDetailsDataTable", and the "CancellationRateByInstitutionLineChart" features.

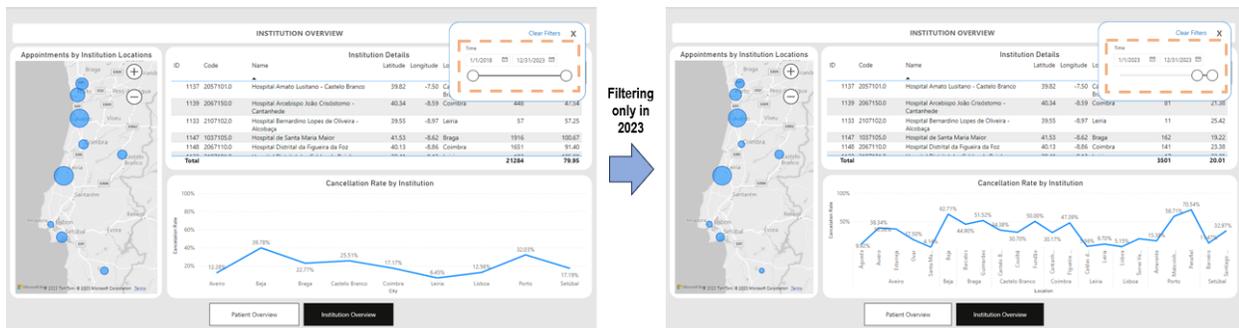

*Figure 9 - The Institution Overview Page shows the data regarding appointments scheduled only in 2023.*

Focusing solely on chart actions, we present the "CancellationRateChart" as an example (cf. Figure 10) to illustrate how the chart adjusts when a user drills down on the data point representing the city of "Porto". This drill-down shifts the chart's focus from displaying cancellation rates by individual institutions to featuring the average cancellation rate across various cities.

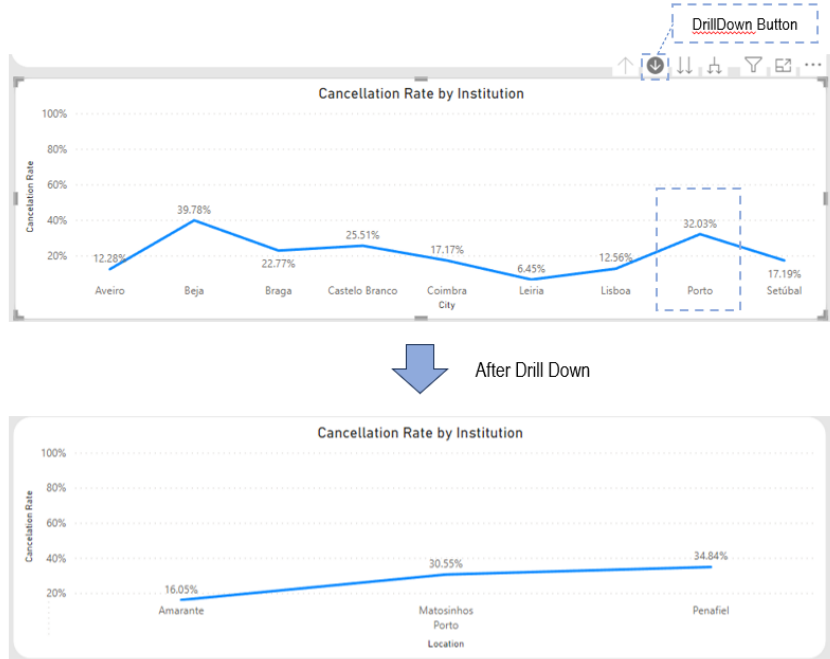

*Figure 10 – The CancellationRateChart appearance after applying a drill-down operation on the city "Porto" (at the bottom)*

The zoom-in and zoom-out functions can be utilised with the "LocationMap" chart, as depicted in Figure 11. This modification in the component's visual is initiated by user interaction, whereby the user zooms into the map to examine the appointment data relevant to "Lisboa".

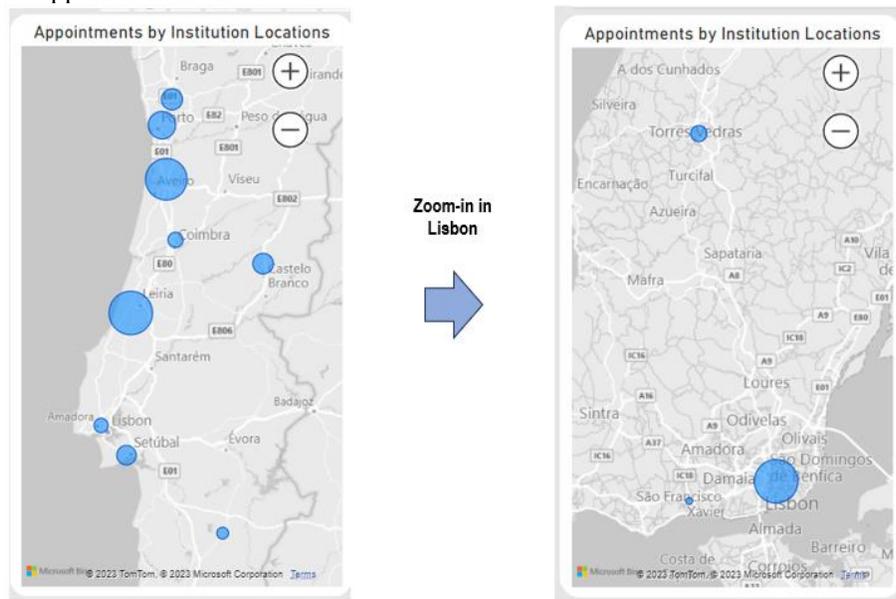

*Figure 11 - The LocationMap appearance after zooming in on Lisbon.*

The two views of the "LocationMap" component, shown in Figure 11, demonstrate the ability of CNLs to precisely delineate the features of current interactive charts, like those presented on PowerBI dashboards and reports. Furthermore, this highlights the possibility of automating the production of such visualisations through both CNL-BI or ASL specifications.

## 8. Final Remarks

Applying CNLs to specifying BI applications, as exemplified by the MEDBuddy-BI illustrative example, showed significant potential for improving the clarity and precision of requirements specifications. This research demonstrates that CNLs can effectively specify requirements for BI applications more systematically and rigorously. Thus, by adopting a structured yet accessible language framework (like the CNL-BI or ASL languages), the inherent ambiguity found in traditional requirements-gathering processes for BI applications can be significantly reduced.

# Apêndice A.   Description and Specification of MEDBuddy-BI

The Appendix A provides a comprehensive description of the MEDBuddy-BI application with complete specifications in CNL-BI and ASL.

MEDBuddy-BI is an analytical NHS add-on constructed with data extracted from the MEDBuddy app. Its primary function is to enhance patient care and facility efficiency by analysing patient interactions and appointment scheduling. This add-on comes equipped with an efficient analytics engine for instantaneous data processing, providing helpful dashboards and reports for healthcare administrators and policymakers. These tools aid in identifying trends, highlighting inefficiencies, and predicting future service demands. The application is currently deployed and can be accessed at [23].

## A.1. Description of the MEDBuddy-BI

The following sections provide an informal overview of the MEDBuddy-BI application and a description of its visual elements.

### A.1.1 Informal description

The informal description of the MEDBuddy-BI application is as following:

```
MEDBuddy is an application for analysing data of the National Health Service. One of the system's functions is
to monitor data on requests for medical consultations in institutions inside and outside the SNS.

A request for a medical appointment has a patient associated with it, the institution where the appointment takes
place, the date the appointment was booked and the date it was completed, and the status of the request, which
can be "booked", "held" or "cancelled", the last two being final statuses. The completion date is the date the
appointment was made, if the appointment was held, or the date the request was cancelled, if the request was
cancelled.
A request for a medical appointment also has a limit on the number of days in which the appointment must take
place after the request enters the "booked" status, known as the maximum response time. It also has associated
with it the number of days that have elapsed between the request moving from the "booked" state to one of the
completion states ("held" or "cancelled"), referred to as the actual response time.

A patient is characterised by their NHS identification number, their gender, age, name, and residence.

A healthcare institution is characterised by its code, name, latitude, longitude, city, and type of institution.
A health institution can be of the Health Centre or Hospital type. A city is characterised by its name, latitude,
and longitude.

The dates on which a medical appointment was booked and when it took place are described by the day, month,
quarter, semester, and year in which they were recorded. This information is especially relevant for analysing
metrics relating to requests for appointments in specific periods, such as the total number of appointments
nationwide in 2022.

The system allows users to interact with it so that they can carry out data analysis operations on the information
relating to requests for medical appointments. There are two types of users of the system:

1. The data scientist at a national level, is responsible for analysing the state of the SNS across the whole
country.
2. The data analyst at a health institution, is responsible for analysing relevant aspects of requests for
consultations for their institution.

Both users can view information regarding medical appointment requests at the Patient and Institution level,
whereas an institution's data analyst will only be able to view this information at the level of their institution.

On the page relating to health institutions, it is possible to view the general information of an institution
(its name, code, city, etc...), as well as the total number of orders, the total number of cancelled orders and
the average time waiting time in each institution, in tabular form. It is also possible to view this information
on a map of Portugal, and the cancellation rate for appointment requests by institution is also displayed.
Finally, the user can also filter the information on this page by location of the institution or group this
information by city.

On the patient page, it is also possible to view general patient information, in a tabular representation, or a
graphical representation, distributed by gender, age, and residence.

On both pages, the user can filter information relating to patients or institutions over some time.
```

## A.1.2 Visual elements description

The description of the visual elements on the Institution Overview Page (Figure 12 and Table 4) and the Patient Overview Page (Figure 13 and Table 5).

**Institution Overview Page**

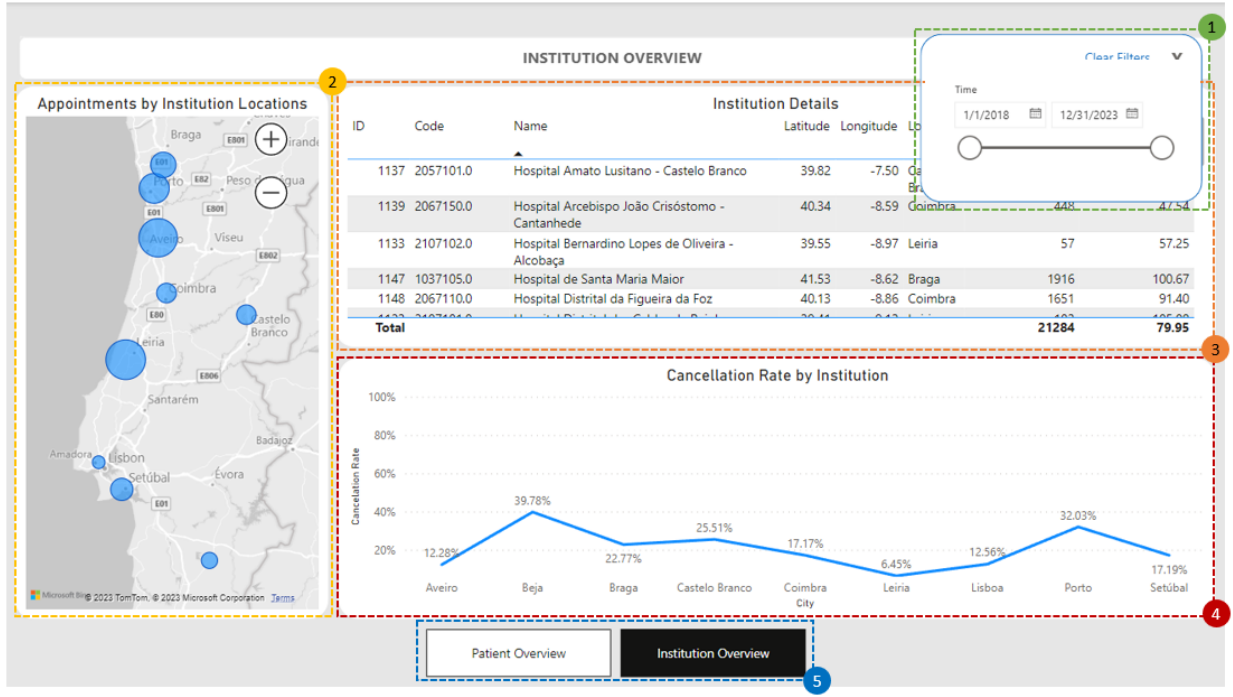

*Figure 12 - Institution Overview Page*

*Table 4 - Institution Overview Page visual elements description*

| Number code on Figure 12 | Name | Type | Description |
|---|---|---|---|
| 1 | TimeRangeFilter | Form | This filter allows users to filter data by specifying a time range, enabling them to view information relevant to a chosen period. |
| 2 | LocationMap | Interactive GeographicalMap | An interactive map that displays geographical data related to healthcare institutions. |
| 3 | InstitutionTable | Table | A comprehensive table that presents information about healthcare institutions. |
| 4 | CancellationRateChart | InteractiveLineChart | An interactive line chart showing cancellation rates over time at various healthcare institutions. |
| 5 | NavigationButtons | Detail | A set of buttons or controls designed to aid users in navigating through the MEDBuddy-BI application. |

## Patient Overview Page

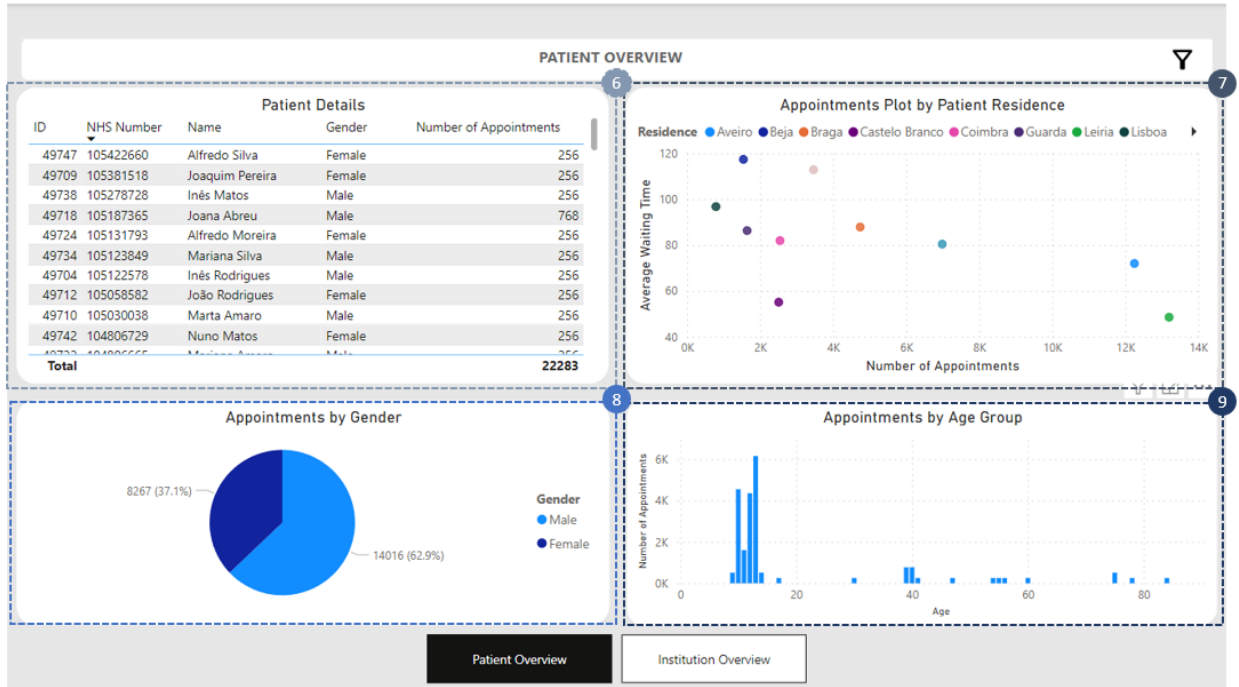

*Figure 13 - Patient Overview Page.*

*Table 5 - Patient Overview Page visual elements description*

| Number code on Figure 13 | Name | Type | Description |
|---|---|---|---|
| 6 | PatientTable | Table | This table displays comprehensive patient data. |
| 7 | PatientAppointment Scatter | InteractiveScatterPlot | An interactive scatter plot visualising the relationship between different variables in patient appointments. |
| 8 | GenderChart | InteractivePieChart | An interactive pie chart that illustrates the distribution of patients by gender. |
| 9 | AgeChart | InteractiveBarChart | An interactive bar chart showing patient distribution across different age groups. |

## A.2. CNL-BI Specification

The following text fragments describe the data model, user interface elements, use cases and OLAP operations represented with the CNL-BI language.

```
A) Multidimensional model

A.1. Data Enumerations:

Data enumeration Gender with values Male and Female.
Data enumeration States with values Booked, Held and Cancelled.
Data enumeration InstitutionTypes with values Health Centre and Hospital.

A.2. Dimensions:

DataEntity Patient is a Master Dimension with attributes
    id is a UUID (PrimaryKey),
    nhs_number is an Integer (NotNull),
    name is a String (NotNull),
    gender is a Gender (NotNull),
    residence refers to Dimension City (NotNull)

DataEntity Institution is a Master Dimension with attributes
    id is a UUID (PrimaryKey),
    code is a String (NotNull),
    name is a String (NotNull),
    latitude is a Decimal (NotNull),
    longitude is a Decimal (NotNull),
    city refers to Dimension City (NotNull),
    type is an InstitutionType (NotNull).

DataEntity City is a Reference Dimension with attributes
    id is a UUID (PrimaryKey),
    latitude is an Decimal (NotNull),
    longitude is an Decimal (NotNull),
    name is a String (NotNull).

DataEntity RequestState is a Reference Dimension with attributes
    id is a UUID (PrimaryKey),
    is_final is a Boolean (NotNull),
    is_initial is a Boolean (NotNull),
    name is a State (NotNull).

DataEntity Time is a Reference Dimension with attributes
    id is a UUID (PrimaryKey),
    date is a Date (NotNull),
    day is an Integer (NotNull),
    month is an Integer (NotNull),
    quarter is an Integer (NotNull),
    semester is an Integer (NotNull),
    year is an Integer (NotNull).

A.3. Facts:

DataEntity AppointmentRequest is a Transactional Fact with attributes
    id is a UUID (PrimaryKey),
    institution refers to Dimension Institution (NotNull),
    patient refers to Dimension Patient (NotNull),
    state refers to Dimension RequestState (NotNull),
    scheduled_date refers to Dimension Time (NotNull),
    closed_date refers to Dimension Time,
    maximum_response_time is an Integer (NotNull),
    actual_response_time is an Integer,
    closed is a Boolean (NotNull),
    CountAppointments is an Integer (operation COUNT(id)),
    CountCancelledAppointments is an Integer (operation COUNT(state = State.Cancelled)),
    CancellationRate is a Decimal (operation (CountCancelledAppointments / CountAppointments)),
    AvgWaitingTime is a Decimal (operation AVERAGE (actual_response_time)),
    MinDate is a Date (operation MIN (scheduled_date)),
    MaxDate is a Date (operation MAX (scheduled_date)).
```

**B) Use Cases**

**B.1. Actors:**

Actor **NationalLevelDataAnalyst** is a User described as a user with permission to visualise the system at a national level.
Actor **InstitutionLevelDataAnalyst** is a User described as a user with permission to visualise the system at an institutional level.

**B.2. Use Cases:**
// Actors
Actor **NationalLevelDataAnalyst** is a User described as a user with permission to visualise the system at a national level.
Actor **InstitutionLevelDataAnalyst** is a User described as a user with permission to visualise the system at an institutional level.

// Use Cases
UseCase **AnalysisAppointmentsInstitutionOnNationalLevel** is a BI_Analysis
  actor NationalLevelDataAnalyst,
  data source AppointmentRequest,
  performs
    // Appointments in a specific year
    OLAP Operation ScheduledAppointmentsInSpecificYear is a Slice
      where Appointment.scheduled_date.year = Time.year
      described as slice the data to visualise the appointments that were scheduled in a specific year,
    // Appointments scheduled in a specific city and year
    OLAP Operation ScheduledAppointmentsBySpecificCityAndYear is a Dice
      where Institution.city = City.id and Appointment.scheduled_date.year = Time.year
      described as dice the data to visualise the appointments that were scheduled in a specific city and year,
    // Appointments by institution's city
    OLAP Operation AppointmentsByInstitutionCity is a Roll-up
      group by Institution.city
      described as rolling up the data to visualise the number of appointments per institution's city,
    // Appointments by institution
    OLAP Operation AppointmentsByInstitution is a Drill-down
      group by Institution.name
      described as rolling up the data to visualise the number of appointments per institution,

  described as it displays the appointment data by institution at a national level.

UseCase **AnalysisAppointmentsInstitutionOnNationalLevel** is a BIAnalysis
  actor InstitutionLevelDataAnalyst,
  data source AppointmentRequest,
  performs
    // Appointments in a specific year
    OLAP Operation ScheduledAppointmentsInSpecificYear is a Slice
      where Appointment.scheduled_date.year = Time.year
      described as slicing the data to visualise the appointments that were scheduled in a specific year,

  described as only displays the appointment data at a single institution level.
// Patient Page Use Cases
UseCase **AnalysisAppointmentsPatientOnNationalLevel** is a BIAnalysis
  actor NationalLevelDataAnalyst,
  data source AppointmentRequest,
  performs
    // Appointments in a specific year
    OLAP Operation ScheduledAppointmentsInSpecificYear is a Slice
      where Appointment.scheduled_date.year = Time.year
      described as slices the data to visualise the appointments that were scheduled in a specific year,
    // Appointments scheduled in a specific year by patients who reside in a specific city
    OLAP Operation ScheduledAppointmentsBySpecificPatientResidenceCityAndYear is a Dice
      where Patient.residence = City.id and Appointment.scheduled_date.year = Time.year
      described as dices the data to visualise the appointments that were scheduled in a specific year by
patients that reside in a specific city,
    // Appointments by patient's gender
    OLAP Operation AppointmentsByGender is a Roll-up
      group by Patient.gender
      described as rolls up the data to visualise the number of appointments per patient's gender,
    // Appointments by patient's age group
    OLAP Operation AppointmentsByAgeGroup is a Roll-up
      group by Patient.age
      described as rolls up the data to visualise the number of appointments per patient's age group,

  described as it displays the appointment data by patient at a national level.

UseCase **AnalysisAppointmentsInstitutionOnNationalLevel** is a BIAnalysis
  actor InstitutionLevelDataAnalyst,
  data source AppointmentRequest,
  performs
    // Appointments in a specific year
    OLAP Operation ScheduledAppointmentsInSpecificYear is a Slice

```
     where Appointment.scheduled_date.year = Time.year
     described as slicing the data to visualise the appointments that were scheduled in a specific year,
  // Appointments scheduled in a specific year by patients who reside in a specific city
  OLAP Operation ScheduledAppointmentsBySpecificPatientResidenceCityAndYear is a Dice
     where Patient.residence = City.id and Appointment.scheduled_date.year = Time.year
     described as dicing the data to visualise the appointments that were scheduled in a specific year by
patients who reside in a specific city,
  // Appointments by patient's gender
  OLAP Operation AppointmentsByGender is a Roll-up
     group by Patient.gender
     described as rolling up the data to visualise the number of appointments per patient's gender,
  // Appointments by patient's age group
  OLAP Operation AppointmentsByAgeGroup is a Roll-up
     group by Patient.age
     described as rolling up the data to visualise the number of appointments per patient's age group,

  described as only displaying the appointment data by a patient at a single institution level.

C) User Interface Elements

// Institution Overview Page
UIContainer InstitutionOverviewPage is a Main Window
  that contains
     // Number 1
     UIComponent TimeRangeFilter is a Form
        data binding to AppointmentRequest,
        starting at AppointmentRequest.MinDate
        and ending at AppointmentRequest.MaxDate,
     // Number 2
     UIComponent LocationMap is an InteractiveGeographicalMap
        data binding to AppointmentRequest
        with latitude Institution.latitude
        longitude Institution.longitude,
        and value AppointmentRequest.CountAppointments,
        actions ZoomAndPanUpdate, DrillDown, TooltipAndHoverDetails
     // Number 3
     UIComponent InstitutionTable is a Table
        data binding to AppointmentRequest
        with columns Institution.id, Institution.code, Institution.name, Institution.latitude,
Institution.longitude, Institution.location, Appointments.CountAppointments, AppointmentRequest.AvgWaitingTime,
     // Number 4
     UIComponent CancellationRateChart is an InteractiveLineChart
        data binding to AppointmentRequest
        with x-axis Intitution.type
        and y-axis AppointmentRequest.CountAppointments,
        actions DrillDown, RealTimeDataUpdate, TooltipAndHoverDetails
     // Number 5
     UIComponent InstitutionPageNavigationButton is a Detail
        that navigates to InstitutionOverviewPage,

     UIComponent PatientPageNavigationButton is a Detail
        that navigates to PatientOverviewPage.

// Patient Overview Page
UIContainer PatientOverviewPage is a Main Window
  that contains
     // Number 6
     UIComponent PatientTable is a Table
        data binding to AppointmentRequest
        with columns Patient.id, Patient.nhs_number, Patient.name, Patient.gender,
AppointmentRequest.CountAppointments,
     // Number 7
     UIComponent PatientAppointmentScatter is a ScatterPlot
        data binding AppointmentRequest
        with x-axis AppointmentRequest.CountAppointments
        y-axis AppointmentRequest.AvgWaitingTime,
        and label Patient.residence,
        actions TooltipAndHoverDetails
     // Number 8
     UIComponent GenderChart is an InteractivePieChart
        data source AppointmentRequest
        with segments defined by Patient.gender,
        and values AppointmentRequest.CountAppointments,
        actions TooltipAndHoverDetails
     // Number 9
     UIComponent AgeBarChart is an InteractiveBarChart
        data source AppointmentRequest
        with x-axis Patient.age
        and y-axis AppointmentRequest.CountAppointments,
```

```
    actions TooltipAndHoverDetails
// Time Range Filter
UIComponent TimeRangeFilter is a Form
  data binding to AppointmentRequest,
  starting at AppointmentRequest.MinDate
  and ending at AppointmentRequest.MaxDate,
// Navigation Buttons
UIComponent InstitutionPageNavigationButton is a Detail that navigates to InstitutionOverviewPage,

  UIComponent PatientPageNavigationButton is a Detail that navigates to PatientOverviewPage.
```

## A.3 ASL Specification

The following text fragments describe the data model, user interface elements, use cases and OLAP operations represented with the ASL language. The complete ASL description of the MEDBuddy-BI can be consulted in [24].

```
A) ASL Types extensions
/* Data Entities SubTypes */
DataEntitySubType BI_Dimension [description "Dimension data entity sub type"]
DataEntitySubType BI_Fact [description "Fact data entity sub type"]

/* Data Entities Attribute Types */
DataAttributeType UUID
DataAttributeType _Dimension [description "Used to reference a Dimension on a data entity attribute"]

/* UIContainers SubTypes */
UIContainerSubType Dashboard
UIContainerSubType Page

/* UIComponents Types */
UIComponentType Card
UIComponentType InteractiveChart
UIComponentType Filter

/*UIComponents Sub Types*/

// Tables
UIComponentSubType Table

// Charts
UIComponentSubType InteractiveLineChart
UIComponentSubType InteractivePieChart
UIComponentSubType InteractiveBarChart
UIComponentSubType InteractiveScatterPlot
UIComponentSubType InteractiveGeographicalMap

// Filters
UIComponentSubType Dropdown
UIComponentSubType Range
UIComponentSubType Search

/* UIComponents Parts Sub Types */

// Tables
UIComponentPartSubType Column

// Charts
UIComponentPartSubType X_Axis
UIComponentPartSubType Y_Axis
UIComponentPartSubType Value
UIComponentPartSubType Area
UIComponentPartSubType Legend
UIComponentPartSubType Label
UIComponentPartSubType Location
UIComponentPartSubType Longitude
UIComponentPartSubType Latitude

// Filters
UIComponentPartSubType Option

/* OLAP Operations */
ActionType BI_Slice [ description "This is a Slice operation"]
ActionType BI_Dice [ description "This is a Dice operation"]
ActionType BI_DrillDown [ description "This is a Drill-down operation"]
ActionType BI_Rollup [ description "This is a Roll-up Operation"]
```

```
ActionType BI_Pivot [ description "This is a Pivot operation"]
/* Use Cases Types */
UseCaseType BI_Analysis [ description "Represents a data analytical use case"]

B) Multidimensional model

B.1. Data Enumerations:

DataEnumeration Gender values (Male, Female)
DataEnumeration States values (Booked, Held, Cancelled)
DataEnumeration InstitutionTypes values (HealthCentre, Hospital)

B.2. Dimensions:

DataEntity e_City "City" : Reference : BI_Dimension [
    attribute id "UUID" : UUID [constraints (PrimaryKey NotNull Unique)]
    attribute latitude "Latitude" : Decimal [constraints (NotNull)]
    attribute longitude "Longitude" : Decimal [constraints (NotNull)]
    attribute Name "Name" : String(50) [constraints (NotNull)]
    description "This dimension represents the details of a city" ]

DataEntity e_Patient "Patient" : Master : BI_Dimension [
    attribute id "UUID" : UUID [constraints (PrimaryKey NotNull Unique)]
    attribute nhs_number "NHS Number" : Integer [constraints (NotNull)]
    attribute age "Age" : Integer [constraints (NotNull)]
    attribute Name "Name" : String(50) [constraints (NotNull)]
    attribute gender "Gender" : DataEnumeration Gender [constraints (NotNull)]
    attribute residence "Residence" : _Dimension [constraints (NotNull ForeignKey(e_City))]
    description "This dimension represents the details of a patient" ]

DataEntity e_Institution "Institution" : Master : BI_Dimension [
    attribute id "UUID" : UUID [constraints (PrimaryKey NotNull Unique)]
    attribute Code "Code" : String(50) [constraints (NotNull)]
    attribute Name "Name" : String(50) [constraints (NotNull)]
    attribute latitude "Latitude" : Decimal [constraints (NotNull)]
    attribute longitude "Longitude" : Decimal [constraints (NotNull)]
    attribute city "City" : _Dimension [constraints (NotNull ForeignKey(e_City))]
    attribute type "Type" : DataEnumeration InstitutionTypes [constraints (NotNull)] ]

DataEntity e_RequestState "Request State" : Reference : BI_Dimension [
    attribute id "UUID" : UUID [constraints (PrimaryKey NotNull Unique)]
    attribute is_final "Is Final" : Boolean [constraints (NotNull)]
    attribute is_initial "Is Initial" : Boolean [constraints (NotNull)]
    attribute Name "Name" : DataEnumeration States [constraints (NotNull)]
    description "This dimension represents the details of a request state" ]

DataEntity e_Time "Time" : Reference : BI_Dimension [
    attribute id "UUID" : UUID [constraints (PrimaryKey NotNull Unique)]
    attribute date "Date" : Date [constraints (NotNull)]
    attribute day "Day" : Integer [constraints (NotNull)]
    attribute month "Month" : Integer [constraints (NotNull)]
    attribute quarter "Quarter" : Integer [constraints (NotNull)]
    attribute year "Year" : Integer [constraints (NotNull)]
    description "This dimension represents the details of the time" ]

B.3. Facts:

DataEntity e_AppointmentRequest "Appointment Request" : Transaction : BI_Fact [
    attribute id "UUID" : UUID [constraints (PrimaryKey NotNull Unique)]
    attribute institution "Institution" : _Dimension [constraints (NotNull ForeignKey(e_Institution))]
    attribute patient "Patient" : _Dimension [constraints (NotNull ForeignKey(e_Patient))]
    attribute _state "Request State" : _Dimension [constraints (NotNull ForeignKey(e_RequestState))]
    attribute scheduled_date "Scheduled Date" : _Dimension [constraints (NotNull ForeignKey(e_Time))]
    attribute closed_date "Closed Date" : _Dimension [constraints (ForeignKey(e_Time))]
    attribute closed "Is closed" : Boolean [constraints (NotNull)]
    attribute maximum_response_time "Maximum Response Time" : Integer [constraints (NotNull)]
    attribute actual_response_time "Actual Response Time" : Integer
    attribute CountAppointments "Count of Appointments" : Integer [formula details: count (e_AppointmentRequest)]
    attribute CountCancelledAppointments "Count of Cancelled Appointments" : Integer [formula details: count
(e_AppointmentRequest) tag (name "expression" value "count(if(was_cancelled = True))")]
    attribute CancellationRate "Cancellation Rate" : Decimal [formula arithmetic (CountCancelledAppointments /
CountAppointments)]
    attribute AvgWaitingTime "Average Waiting Time" : Decimal [tag (name "expression" value
"average(actual_waiting_time)")]
    attribute MinDate "Minimum Date" : Date [tag (name "expression" value "min(time.date)")]
    attribute MaxDate "Maximum Date" : Date [tag (name "expression" value "max(time.date)")] ]

B.4. Data Entities Clusters:
```

```
DataEntityCluster dec_Appointments "Appointments Cluster" : Transaction [
  main e_AppointmentRequest
  uses e_Time, e_Patient, e_Institution, e_City
  description "Data entity cluster to represent the connections between the Appointments fact with the
dimension" ]

C) Use Cases

C.1. Actors:

Actor a_National_Level_Data_Analyst "National Level Data Analyst" : User [ description "User with permitions to
visualise the application at a national level" ]
Actor a_Institution_Manager "Institution Manager" : User [ description "User with permitions to visualise the
application at an intitution level" ]

C.2. Use Cases:

// Institution Page Use Cases
UseCase uc_Analysis_Appointments_National_Level : BI_Analysis [
  actorInitiates a_National_Level_Data_Analyst
  dataEntity dec_AppointmentRequests

  actions BI_Slice, BI_Dice, BI_Rollup, BI_DrillDown, BI_Pivot

  tag (name "BI-Action:BI_Slice:ScheduledAppointmentsInSpecificYear" value "Dimensions:'e_Time'")
  tag (name "BI-Action:BI_Dice:ScheduledAppointmentsBySpecificCityAndYear" value "Dimensions:'e_Time,
e_Institution, e_City'")
  tag (name "BI-Action:BI_RollUp:AppointmentsByInstitutionCity" value "Dimensions:'e_Institution'")
  tag (name "BI-Action:BI_RollUp:AppointmentsByInstitution" value "Dimensions:'e_Institution'")

  description "Displays the appointment data by institution at a national level" ]

UseCase uc_Analysis_Appointments_Institution_Level : BI_Analysis [
  actorInitiates a_Institution_Level_Data_Analyst
  dataEntity dec_AppointmentRequests

  actions BI_Slice, BI_Dice, BI_Rollup, BI_DrillDown, BI_Pivot

  tag (name "BI-Action:BI_Slice:ScheduledAppointmentsInSpecificYear" value "Dimensions:'e_Time'")

  description "Only displays the appointment data at a single institution level" ]

// Patient Page Use Cases
UseCase uc_AnalysisAppointmentsPatientOnNationalLevel : BI_Analysis [
  actorInitiates a_National_Level_Data_Analyst
  dataEntity dec_AppointmentRequests

  actions BI_Slice, BI_Dice, BI_Rollup, BI_DrillDown, BI_Pivot

  tag (name "BI-Action:BI_Slice:ScheduledAppointmentsInSpecificYear" value "Dimensions:'e_Time'")
  tag (name "BI-Action:BI_Dice:ScheduledAppointmentsBySpecificPatientResidenceCityAndYear" value
"Dimensions:'e_Time, e_Patient, e_City'")
  tag (name "BI-Action:BI_RollUp:AppointmentsByGender" value "Dimensions:'e_Patient'")
  tag (name "BI-Action:BI_RollUp:AppointmentsByAge" value "Dimensions:'e_Patient'")

  description "Displays the appointment data by patient at a national level" ]

UseCase uc_AnalysisAppointmentsInstitutionOnNationalLevel : BI_Analysis [
  actorInitiates a_Institution_Level_Data_Analyst
  dataEntity dec_AppointmentRequests

  actions BI_Slice, BI_Dice, BI_Rollup, BI_DrillDown, BI_Pivot

  tag (name "BI-Action:BI_Slice:ScheduledAppointmentsInSpecificYear" value "Dimensions:'e_Time'")
  tag (name "BI-Action:BI_Dice:ScheduledAppointmentsBySpecificPatientResidenceCityAndYear" value
"Dimensions:'e_Time, e_Patient, e_City'")
  tag (name "BI-Action:BI_RollUp:AppointmentsByGender" value "Dimensions:'e_Patient'")
  tag (name "BI-Action:BI_RollUp:AppointmentsByAge" value "Dimensions:'e_Patient'")

  description "Only displays the appointment data by patient at a single institution level" ]
```

**D) User Interface specification**

**D.1. Filters:**
// Time Range Filter
component uiCo_TimeFilter "Select Time Range" : Filter : Range [
    dataBinding dec_AppointmentRequests

    part minDate "Min Date" : Field : Option [dataAttributeBinding e_AppointmentRequest.MinDate]
    part maxDate "Max Date" : Field : Option [dataAttributeBinding e_AppointmentRequest.MaxDate]
    // Submit Button
    event e_ApplyFilter : Submit : Submit_Update [tag (name "Time Slice" value "Slice Appointments by the
selected time range")]
    event e_ResetFilter : Submit : Submit_Update [tag (name "Reset" value "Clear the time filter")] ]

**D.2. Tables:**
// Patient Table
component uiCo_PatientTable "Patient Details" : List : Table [
    dataBinding dec_AppointmentRequests
    // The data table columns
    part id "ID" : Field : Column [dataAttributeBinding e_Patient.id]
    part nhs_number "NHS Number" : Field : Column [dataAttributeBinding e_Patient.nhs_number]
    part Name "Name" : Field : Column [dataAttributeBinding e_Patient.Name]
    part gender "Gender" : Field : Column [dataAttributeBinding e_Patient.gender]
    part CountAppointments "Number of Appointments" : Field : Column [dataAttributeBinding
e_AppointmentRequest.CountAppointments] ]

// Institution Table
component uiCo_InstitutionTable "Institution Details" : List : Table [
    dataBinding dec_AppointmentRequests
    // The table columns
    part id "ID" : Field : Column [dataAttributeBinding e_Institution.id]
    part Code "Code" : Field : Column [dataAttributeBinding e_Institution.Code]
    part Name "Name" : Field : Column [dataAttributeBinding e_Institution.Name]
    part latitude "Latitude" : Field : Column [dataAttributeBinding e_Institution.latitude]
    part longitude "Longitude" : Field : Column [dataAttributeBinding e_Institution.longitude]
    part city "City" : Field : Column [dataAttributeBinding e_Institution.city]
    part AvgWaitingTime "Average Waiting Time" : Field : Column [dataAttributeBinding
e_AppointmentRequest.AvgWaitingTime] ]

**D.3. Charts:**
// Bar Chart
component uiCo_AgeBarChart "Age Distribution" : InteractiveChart : InteractiveBarChart [
    dataBinding dec_AppointmentRequests
    // The chart axes
    part age "Age" : Field : X_Axis [dataAttributeBinding e_Patient.age]
    part CountAppointments "Number of Appointments" : Field : Y_Axis [dataAttributeBinding
e_AppointmentRequest.CountAppointments]

    event TooltipAndHoverDetails : Other ]

// Pie Chart
component uiCo_GenderChart "Gender Distribution" : InteractiveChart : InteractivePieChart [
    dataBinding dec_AppointmentRequests
    // The chart segments
    part gender "Gender" : Field : Label [dataAttributeBinding e_Patient.gender]
    part numberOfAppointments "Number of Appointments" : Field : Value [dataAttributeBinding
e_AppointmentRequest.CountAppointments]

    event TooltipAndHoverDetails : Other ]

// Line Chart
component uiCo_CancellationRateChart "Cancellation Rate by Institution" : InteractiveChart :
InteractiveLineChart [
    dataBinding dec_AppointmentRequests
    // The chart axes
    part type "Institution Type" : Field : X_Axis [dataAttributeBinding e_Institution.type]
    part CountAppointments "Number of Appointments" : Field : Y_Axis [dataAttributeBinding
e_AppointmentRequest.CountAppointments]

    event DrillDown : Submit : Submit_Update [navigationFlowTo uiCo_CancellationRateChart]
    event RealTimeDataUpdate : Submit : Submit_Update [navigationFlowTo uiCo_CancellationRateChart]
    event TooltipAndHoverDetails : Other ]

```
// Scatter Plot
component uiCo_PatientAppointmentScatter "Appointment Scatter Plot by Patient Residence" : InteractiveChart :
InteractiveScatterPlot [
  dataBinding dec_AppointmentRequests
  // The chart axes
  part CountAppointments "Number of Appointments" : Field : X_Axis [dataAttributeBinding
e_AppointmentRequest.CountAppointments]
  part AvgWaitingTime "Average Waiting Time" : Field : Y_Axis [dataAttributeBinding
e_AppointmentRequest.AvgWaitingTime]
  // The label
  part residence "Patient Residence" : Field : Label [dataAttributeBinding e_Patient.residence]

  event TooltipAndHoverDetails : Other ]

// Location Map
component uiCo_LocationMap "Appointments by Institution Locations" : InteractiveChart :
InteractiveGeographicalMap [
  dataBinding dec_AppointmentRequests
  // The chart axes
  part latitude "Latitude" : Field : Latitude [dataAttributeBinding e_Institution.latitude]
  part longitude "Longitude" : Field : Longitude [dataAttributeBinding e_Institution.longitude]
  // The value
  part CountAppointments "Number of Appointments" : Field : Value [dataAttributeBinding
e_AppointmentRequest.CountAppointments]

  event ZoomAndPanUpdate : Submit : Submit_Update [navigationFlowTo uiCo_CancellationRateChart]
  event RealTimeDataUpdate : Submit : Submit_Update [navigationFlowTo uiCo_CancellationRateChart]
  event TooltipAndHoverDetails : Other ]

D.4. Pages:

// Patient Overview Page
UIContainer uiCo_PatientOverviewPage "Patient Overview Page" : Window : Page [
  // Visual elements
  component uiCo_PatientTable
  component uiCo_GenderChart
  component uiCo_AgeBarChart
  component uiCo_PatientAppointmentScatter
  // Filters
  component uiCo_TimeFilter
  // Buttons
  event ev_InstitutionPageButton : Submit : Submit_Back [navigationFlowTo uiCo_InstitutionOverviewPage]
]

// Institution Overview Page
UIContainer uiCo_InstitutionOverviewPage "Institution Overview Page" : Window : Page [
  // Visual elements
  component uiCo_InstitutionTable
  component uiCo_CancellationRateChart
  component uiCo_LocationMap
  // Filters
  component uiCo_TimeFilter
  // Buttons
  event ev_PatientPageButton : Submit : Submit_Back [navigationFlowTo uiCo_PatientOverviewPage]
]
```